# Collapse of Point Vortex Dipoles in a Bounded Fluid Layer and the Hydrodynamic Mechanism of Mutual Attraction of Like-Charged Micro-Particles in the Colloid or Dusty Plasma Systems

S.G.Chefranov[1], A.G. Chefranov[2]

[1] A.M. Obukhov Institute of Atmospheric Physics RAS, Moscow  schefranov@mail.ru

[2] Eastern Mediterranean University, Famagusta, North Cyprus  Alexander.chefranov@emu.edu.tr

**Abstract**

The new exact weak solution of the equations for the ideal incompressible fluid dynamics in the finite layer inside two plates with solid boundaries is obtained. The solution meets a devised non-linear finite dimension Hamiltonian dynamic system for the coordinates and the Lamb impulses of N point vortex dipoles (PVD), i.e. extremely small solid spherical particles, moving with respect to the fluid. For N=2 the necessary condition for the collapse (or the converging in one point during the finite time) of two PVD is stated. On the base of the proposed theory, the new hydrodynamic mechanism of converging for two small spherical particles (of the same radius) is introduced and used for interpreting the observed paradoxical effects of attraction for micro particles with the same sign of electrical charge in the colloid and dusty plasma systems. The correspondence of the condition for collapse of two PVD with the experimentally observed data where convergence of two like-charged micro-spheres moving in the fluid layer is observed only when definite necessary restrictions on the fluid layer depth are imposed from above and below (J. C. Crocker, D. G. Grier, 1996) is stated. We also state the possibility of correspondence between the obtained condition of collapse for two PVD in the limit of unbounded fluid and known numerical results on the threshold of the impact parameter for the realization of resonance bond state (and corresponding stochastic dynamic regime) for two finite size vortex dipoles during their scattering in unbounded fluid (S. V. Manakov, L. N. Shchur, 1983; L. Tophoj, H. Aref, 2008). Here we show that in the latter process the size of dipoles is not so important as it was thought before.

*Introduction*

Two-phase systems in the form of micro-particles moving in the liquid, gas, or plasma medium, are widely met in natural and artificial systems; it makes them an important object for fundamental and applied research [1-11]. Meanwhile, a problem of defining the leading physical mechanisms of forming meta-stable structures from like-charged colloid or dusty plasma micro-particles still is not solved. The particles usually have micron sizes and large negative charge not preventing however forming of respective colloid and dusty plasma crystals [1, 5]. Related with it, a paradoxical effect of arising of the strong long-range mutual attraction between two like-charged micro-spheres is experimentally found [1, 2] (see also [3]). So, in [1], the effect of attraction is observed for two like-charged latex micro-spheres placed in a plane thin water layer bounded by two solid smooth glass surfaces having the same sign of electric charge as that of the micro-sphere particles. It was found out that the presence of the fluid layer solid boundaries may play an important role for realization of the effect of mutual attraction under defined in [1] constraints on the layer depth from above and below. Constraint on the layer depth from above as it was shown in [2, 7] is not so unexplainable as the constraint from below that up to now has not definite physical explanation.

Indeed there is no commonly accepted physical explanation of the cause of emerging of the observed long-range strong mutual attraction between like-charged micro-spheres in the medium. Usually, as an explanation, non-equilibrium processes are considered related with redistribution of charged microscopic medium particles surrounding the both interacting micro-spheres and leading to "the shadow force effects" of their mutual attraction [3, 5]. From the other side, in [8-10], it is proposed a non-equilibrium hydrodynamic mechanism, caused by the macroscopic motion of the both micro-spheres with respect to the medium. Similar hydrodynamic mechanism is considered also in [12-15] in relation with the effect of coagulation of aerosol and fog particles in an acoustic field that has practical importance for the opportunity of artificial falling of such particles from the atmospheric air.

In the present work, we propose a generalization of the theory [16] for the case of the bounded fluid layer where it may be effectively realized a hydrodynamic mechanism of mutual attraction of two micro-spheres even without taking into account viscosity medium effects. This is the main distinction of the suggested theory from [8-10, 12, 17-19] where just viscous effects are considered as the main cause for emerging of hydrodynamic interaction between moving in the medium micro-particles. Non-viscous potential flow providing hydrodynamic interaction of the moving in the fluid solid spherical particles was considered also in [13], but there were obtained too small velocities of micro-particles converging compared to the observation data and the next development of this research direction in [12] was made by counting only the viscous effects. Instead of viscosity considering, we propose in the frame of the developed in the present paper theory, a new model of the strong micro-particles interaction on the base of the exact weak solution of the two dimensional Helmholtz equation (see in Appendix, (A.8)-(A.11)) for vortex conservation; the solution is obtained from the derived non-linear dynamic Hamiltonian system for N point vortex dipoles (PVD). In this dynamic system the canonic conjugated variables are the PVD coordinates and Lamb's impulses. At the same time, even when using in [18,19] of hydrodynamic objects converting to PVD in the limit of small particles radii, consistent description of time dependence of the coordinates and velocities (defining micro-particle Lamb's impulses) is absent because in [18, 19], micro-particle velocities with respect to the medium are only fixed by initial date.

Our theory in the limit of unbounded fluid also coincides with the theory proposed in [16] and may be used, as [16], to give the new interpretation for the numerical experiments data obtained for the unbounded fluid for scattering of two finite size vortex dipoles or two point vortex pairs (see the Conclusion and Discussion at the end of this paper).

## 1. Hamiltonian Dynamic Equations for $N \geq 2$ PVD

1. A spherical particle, having radius $R$ and moving with the constant speed $\mathbf{V}$, creates in the surrounding ideal incompressible fluid (on sufficiently large distance) a velocity field coinciding with the velocity field created by a PVD [16, 20]. In the two-dimensional case considered in the present paper, the vortex field for a system of $N$ PVD has the following form:

$$\omega = \varepsilon_{ij} \sum_{\alpha=1}^{N} \gamma_j^{(\alpha)} \frac{\partial \delta(\mathbf{x} - \mathbf{x}_\alpha)}{\partial x_i}, \qquad (1.1)$$

where $\varepsilon_{11} = \varepsilon_{22} = 0, \varepsilon_{12} = -\varepsilon_{21} = 1, i = 1,2, j = 1,2, \gamma_i^{(\alpha)}(t) = \lim\limits_{R_\alpha \to 0, |V_\alpha| \to \infty} P_i^{(\alpha)}; P_i^{(\alpha)} = \pi R_\alpha^2 V_i^{(\alpha)}$ is the Lamb impulse for the PVD with the number $\alpha$, located at the point with coordinates $\mathbf{x}_\alpha(t)$ depending on time $t$. Here, $R_\alpha$ is the radius and $V_i^{(\alpha)}$ is the velocity of a spherical particle with the number $\alpha = 1, \ldots, N$, and it is assumed existence of the finite limit for the values of the Lamb impulses, when the radius of the solid spherical particle tends to zero (we consider here $R_\alpha = R$ for all $\alpha = 1,..,N$). The presentation (1.1) for two-dimensional case is obtained from the three dimensional formula for PVD having the following form: $\omega_i = \varepsilon_{ijl} \sum_{\alpha=1}^{N} \gamma_l^{(\alpha)} \frac{\partial \delta(\vec{x} - \vec{x_\alpha})}{\partial x_j}$ where i=3. The velocity field created in the fluid by three dimensional PVD (extremely small vortex ring, or spherical Hill's vortex) outside the point of its localization is the same as the velocity field produced in the fluid by a moving sphere (in the three dimensional case the Lamb impulse is represented as $P_i^{(\alpha)} = 2\pi R_{(\alpha)}^3 V_i^{(\alpha)}$, i=1,2,3; this determination is different from the used above formula for the two–dimensional case, where R is the radius of solid ring in plane, when for three dimensional case R is the radius of solid sphere).

The distribution of the vortex field for N PVD in the unbounded two-dimensional space corresponds to the stream function (defined as the solution of the following equation $\Delta_2 \Psi = -\omega$, where $\Delta_2$ is the two-dimensional Laplace operator) and velocity field potential $\Phi$ having the following form [16, 20]:

$$\Psi = \sum_{\alpha=1}^{N} \Psi_\alpha, \Psi_\alpha = -\frac{\varepsilon_{ij} \gamma_j^\alpha (x_i - x_i^\alpha)}{2\pi |\mathbf{x} - \mathbf{x}_\alpha|^2}; \Phi = \sum_{\alpha=1}^{N} \Phi_\alpha, \Phi_\alpha = -\frac{\gamma_i^\alpha (x_i - x_i^\alpha)}{2\pi |\mathbf{x} - \mathbf{x}_\alpha|^2}. \qquad (1)$$

2. When the solid plane boundaries for the fluid layer with *N* PVD exist, boundary conditions of the mass non-penetrability through the fluid rigid boundaries shall be met. They may be satisfied by mapping each of the considered PVD on the infinite set of its mirror images. If we select the reference frame at the middle of the fluid layer and direct the *y* axis perpendicular to the layer boundary, and the *x* axis along the layer boundary, then the mirror images of each PVD have one and the same value of the projection of the Lamb impulse on the axis *x* as the PVD located inside the fluid layer. At the same time, for the projection on the *y* axis, it is required only sign change with respect to the corresponding projection of the Lamb impulse vector for the pre-image. Such a system of PVD and all their mirror images now can be considered in the unbounded space for which in [16] in general form (including also three-dimensional case), there was obtained an exact weak solution of the Helmholtz equation for the vortex field. Actually the problem of getting the finite-dimensional dynamic Hamiltonian system for PVD (solution of which is an exact weak solution of the Helmholtz equation) in the case of the bounded fluid layer is reduced to the generalization of given in [16] conclusions to the velocity field potential created in the bounded fluid layer by each of *N* considered particles (PVD) and the infinite set of their mirror images.

To get velocity field potential for the system of *N* PVD located inside the bounded by plane boundaries fluid layer, we extend (1) by adding, for each of *N* PVD, contributions of all its mirror images set of which is infinite. The set of mirror images of a PVD consists, in general case, of two subsets that leads to existence of two terms in the brackets in the obtained in Appendix (from (A.1)) formula (2) (see also illustrating it Fig.3 in [18]) for the complex potential $W = \Phi + i\Psi$:

$$W = -\frac{1}{4H}\sum_{\alpha=1}^{N}\left[\Gamma_\alpha cth\frac{\pi(z-z_\alpha)}{2H} + \Gamma_\alpha^* th\frac{\pi(z-z_\alpha^*)}{2H}\right]; z = x+iy, z_\alpha = x_\alpha + iy_\alpha, \Gamma_\alpha = \gamma_x^\alpha + i\gamma_y^\alpha,  \quad (2)$$

where asterisk denotes complex conjugation of the corresponding values in (2). The expression (2) generalizes the formula (1) in [18] for the spherical particles potential representation obtained in the case of non-zero orthogonal to the boundaries Lamb impulse components. Thus (2) exactly coincides with formula (1) in [18] only when $x_\alpha = 0, \gamma_y^\alpha = 0$ in (2) and when the limit of small micro-sphere radii is realized (and also in the case of absence of the fluid flow at the infinity - in [18], the condition of absence of fluid flow at infinity corresponds to equality to zero of the value of the oil velocity $u_{oil}^\infty = 0$).

3. We propose the model of dynamic interaction for colloid micro-spheres on the base of Hamiltonian equations describing mutual PVD dynamics, which correspond to the weak exact solution of the Helmholz equation for ideal incompressible fluid, considered in [16], by taking into account representation (2) for the fluid layer of the finite depth. In the result, we have the following modified equations for $N \geq 2$ PVD ($\beta = 1,...,N$):

$$\frac{d\gamma_j^{(\beta)}}{dt} = -\gamma_l^{(\beta)} \frac{\partial^2 \Phi(\mathbf{x}_{(\beta)})}{\partial x_j^{(\beta)} \partial x_l^{(\beta)}} = -\frac{\partial E}{\partial x_j^{(\beta)}}$$

$$\frac{dx_j^{(\beta)}}{dt} = \frac{\partial \Phi(\mathbf{x}_{(\beta)})}{\partial x_j^{(\beta)}} = \frac{\partial E}{\partial \gamma_j^{(\beta)}} \tag{3}$$

$$E = \sum_{\beta=1}^{N} \gamma_j^{(\beta)} \frac{\partial \Phi}{\partial x_j^{(\beta)}} = \int d^2 x \, \omega \Psi$$

where contrary to [16], the velocity field potential $\Phi$ is defined now by the new representation (2). In (3), the value of Hamiltonian $E$ corresponds to the interaction energy for $N \geq 2$ PVD in the fluid layer of the finite depth $H$ (see also (A.1), (A.2) in Appendix). For the Hamiltonian $E$ of the dynamic system (3), canonical conjugated variables are the PVD coordinates and Lamb's impulses. Contrary to the case of unbounded fluid, translational invariance of the Hamiltonian takes place in the general case only along the $x$ axis. Respectively, in addition to $E$, the Lamb impulse sum x-component is also invariant: $P_x = \sum_{\alpha=1}^{N} \gamma_x^\alpha = const$. In the Appendix, however, the conditions (A.3)-(A.7) are given that provide for $N = 2$ invariance of $P_y = \gamma_{1y} + \gamma_{2y}$ also, which corresponds to the invariance of "center of gravity" position $R_x = const; R_y = const$ (with $R_x = (x_1 + x_2)/2 = 0; R_y = (y_1 + y_2)/2 = 0$) for this PVD system.

## 2. Solution for Two PVD (N=2)

1. For the case corresponding to these both summed Lamb impulse invariance conditions from (A.7), it is not difficult to get in the limit of $l(0)/H \ll 1$ ($l(0)$ is the initial distance between the PVD - or particles) the system of equations for the PVD coordinates and Lamb impulses for two PVD (with N=2 in (2), (3), see also (A.7)). In the case of $H \to \infty$ this system exactly coincides with the result of [16] in two dimensional fluid (it must be also taken into account that here we use the representation $\mathbf{\gamma} = \mathbf{\gamma}_{(2)} = -\mathbf{\gamma}_{(1)}$ whereas in [16], inversely, $\mathbf{\gamma} = \mathbf{\gamma}_{(1)} = -\mathbf{\gamma}_{(2)}$ with the same notation for the vector distance between two PVD: $\mathbf{l} = \mathbf{x}_{(1)} - \mathbf{x}_{(2)}$). Thus the dynamic system for two particles (PVD) in limit $l(0)/H \ll 1$ has the following form (with accuracy up to including the terms of the order of $l^2/H^2$):

$$\frac{dl}{dt} = \frac{\gamma \cos \psi}{\pi l^2}\left(1 - \frac{\pi^2 l^2 (\cos \varphi \cos \varphi_1 + \cos \psi)}{6H^2 \cos \psi}\right)$$

$$\frac{d\varphi}{dt} = \frac{\gamma \sin \psi}{\pi l^3}\left(1 + \frac{\pi^2 l^2 (\sin \varphi \cos \varphi_1 + \sin \psi)}{6H^2 \sin \psi}\right) \tag{4}$$

$$\frac{d\gamma}{dt} = -\frac{\gamma^2 \cos\psi(3 - 4\cos^2\psi)}{\pi d^3} + O(\frac{l^4}{H^4})$$

$$\frac{d\varphi_1}{dt} = -\frac{\gamma \sin\psi(1 - 4\cos^2\psi)}{\pi d^3} + O(\frac{l^4}{H^4}), \tag{5}$$

where $l_x = l\cos\varphi, l_y = l\sin\varphi, \gamma_x = \gamma\cos\varphi_1, \gamma_y = \gamma\sin\varphi_1, \psi = \varphi - \varphi_1$.

System (4), (5) has the following invariant

$$E = \frac{\gamma^2}{\pi d^2}(1 - 2\cos^2\psi) + \frac{\pi^2 l^2}{6H^2}(1 + \cos^2\varphi_1)). \tag{6}$$

To obtain the solution of (4), (5) in the limit $q = l/H \to 0$ (see also [16]) we may use that from (4), (5) it is possible to state: $\frac{d(l\gamma\cos\psi)}{dt} = -2E = const$. The solution of (4), (5) for q=0 is:

$$l(t) = l(0)\left[1 - \frac{4(Et^2 - t\gamma(0)l(0)\cos\psi(0))}{\pi d^4(0)}\right]^{1/4} \tag{6.1}$$

Thus from (6.1) it is obvious that under condition, when E in (6) for q=l/H=0 is positive (for example when $\psi(0) = \pi/2$), the collapse of two PVD is possible in the finite time $t = t_c = l^2(0)\pi^{1/2}/2E^{1/2} \cong l^3(0)/2\pi^{1/2}R^2V$ in the limit of infinitely large layer depth H (see also [16]). For the time $t > t_c$ the solution (6.1) is not determined because for this times the solution is imaginary and has not real physical value. This means that two particles (PVD) for $t > t_c$ create the bound resonant long–live state when $l(t) \to 0$ at $t \to t_c$ according to (6.1).

To obtain the estimation of $t_c$, let's assume according to [22], that l(0)=16 micron, R=4 micron, V=0.5 sm/sec. Then $t_c \approx 0.01$ sec, that corresponds to the opportunity of effective acoustic dust (or suspended in the air fog drops) coagulation even without taking into account molecular medium viscosity. At the same time, in [13], for the time of the particles converging, there was obtained the value of 1.75 hour that led to the wrong, as it is found out now, conclusion on ineffectiveness of hydrodynamic particles interaction mechanism without counting the medium molecular viscosity (see [12]). Thus, the very presence of strong PVD hydrodynamic interaction described by the Hamiltonian dynamic system (3), as it is seen, may be, without viscosity counting, successfully used when modeling the processes of acoustic aerosol coagulation and colloid and dust-plasma crystals forming.

2. From (3) (and (A.7)), in the other limit case with *l(0)/H>>1* for *N=2*, we have

$$dl_x/dt = \frac{2\pi\gamma}{H^2}\exp(-\frac{\pi l_x}{H})\sin(\varphi_1 - \pi l_y/2H)\sin(\pi l_y/2H),$$
$$dl_y/dt = -dl_x/dt\, ctg(\pi l_y/2H),$$
$$d\gamma_x/dt = -\pi E/2H = const, \quad (7)$$
$$d\gamma_y/dt = d\gamma_x/dt\, ctg(\varphi_1 - \pi l_y/2H),$$
$$E = \frac{\pi\gamma^2}{H^2}\exp(-\pi l_x/H)(1-\cos(2\varphi_1 - \pi l_y/H)) = const$$

From (7) it follows that in the limit of l(0)/H>>1 the collapse of two PVD to a single point is already realizable only when the additional invariant $L = \frac{l_x\pi}{2H} - \ln(\cos(\frac{l_y\pi}{2H})) = L_0 = const$, following from the second equation in (7), is zero (when $L_0 = 0$ for all initial conditions when the two PVD have different not coinciding with each other initial coordinates). For the initial conditions when $L_0 = 0$ from the first equation of (7) and the determination of positive energy E in (7) we may obtain the next exact solution of (7) corresponding to the collapse regime for two particles (PVD) in the finite time $t = t_{c0}$:

$$l_y(t) = l_y(0)(1 - \frac{t}{t_{c0}});$$
$$l_x(t) = \frac{2H}{\pi}\ln\left|\cos(\frac{\pi l_y(0)}{2H}(1-\frac{t}{t_{c0}}))\right|; t_{c0} = \frac{l_y(0)H}{\sqrt{2\pi E}} \quad (8)$$

In (8) the time of collapse $t_{c0}$ in the considered limit of q= l(0)/H>>1 may have very large values as due to this limit q>>1, so to the very rapid decreasing of energy E with increasing of q (see Fig. 2 below). Thus a conclusion on the arising of collapse for two PVD in the limit of arbitrary small fluid layer depths for so large values of collapsing time (that in any real experiment can't be detected in the finite time of experiment continuation) may also qualitatively correspond to the obtained in [1] conclusion on the necessity of layer depth restrictions from below for the opportunity of convergence of two like-charged micro-particles.

Contrary to (6.1), the solution (8) is real also for the time $t > t_{c0}$ when two particles (PVD) continue their motion after collapse and the time of life for the corresponding resonant state is zero $T_{res} = 0$ for (8) and infinite for (6.1), where $T_{res} \to \infty$.

3. On the base of consideration of the direct numerical solutions of equations (3) for $N = 2$, we show (see Fig.1) that for q=l(0)/H < 0.1 the condition of collapse for two particles (PVD) actually numerical coincides with the condition E>0 in (6) for l/H=0 (that is also obtained in [16] in the two-dimensional case for the unbounded fluid). From the other side, for 1.4<q<2 (see Fig. 1), collapse can take place in more wide cases than in the unbounded fluid. Actually, for q from this narrow interval, collapse of two PVD may take place for any initial angle $\psi$ value (between the vectors $\mathbf{l} = \mathbf{x}_1 - \mathbf{x}_2$ and $\mathbf{\gamma} = \mathbf{\gamma}_2 = -\mathbf{\gamma}_1$). It essentially differs from the collapse conditions considered in [16] for the unbounded space and for condition E>0 in (6), where l/H=0. Actually, according to [16] and to condition (6) in the limit l/H=0, collapse is possible only for the angles defined from the inequality $|\cos(\psi(0))| < 1/\sqrt{2}$, i.e. for $45^0 < \psi(0) < 135^0$. At the same time, presence of boundaries results in the arising of additional to $\psi(0)$ controlling parameter q=l(0)/H, defining the collapse possibility for two particles (PVD). Meanwhile, it is found out that even for $\psi(0) \approx \pi$ (i.e. for the case when according to [16] the particle must definitely scatter in the absence of boundaries when E<0 in (6) for l/H=0), collapse is still possible for not too small values of the parameter q (see Fig.1).

Thus, obtained in the experiments [1] restrictions on the layer depth from above can get pure hydrodynamic explanation for the fixed value of initial distance l(0) (when from Fig.1 the collapse take place for all H<l(0)/2). Let us note that considered here non-equilibrium hydrodynamic mechanism of mutual attraction of like-charged particles does not depend explicitly on the specifics of the medium and can manifest itself as for the interaction of aerosol and colloid particles, so for the dusty particles in plasma.

4. In the general case, energy of two PVD for symmetric initial conditions $y_1 + y_2 = 0, \mathbf{\gamma} = \mathbf{\gamma}_2 = -\mathbf{\gamma}_1$ has the following form:

$$E = \frac{\pi\gamma^2}{2H^2}\left[\frac{1}{ch\frac{\pi l \cos\varphi}{H}+1} - \frac{a\cos 2\varphi_1 + b\sin 2\varphi_1}{a^2+b^2}\right], \quad (9)$$

$$a = ch\frac{\pi l \cos\varphi}{H}\cos\frac{\pi l \sin\varphi}{H} - 1, b = sh\frac{\pi l \cos\varphi}{H}\sin\frac{\pi l \sin\varphi}{H}$$

In (9) $\varphi_1 = \varphi - \psi$. Expression (1) in the limit of $q = l/H \ll 1$ coincides with (6).

In particular, for q<1, it is possible to consider the case when in (9): $\varphi = \pi/2$. Then b=0 and energy (9) is positive when

$$\cos^2\psi < (1+\sin^2 \pi l/2H)/2 \quad (10)$$

For example, for q=l/H=0.5, threshold values of the angle $\psi$ have the form $30^0 < \psi < 150^0$ derived from (10) instead of $45^0 < \psi < 135^0$ for q=l/H=0 in (10). If in (10) $q = l/H \to 1+2m, m=1,2,...$ the restriction on $\psi$ have the representation: $0^0 < \psi < 180^0$.

An exact analytical form for the dependency of the threshold angle $\psi$ on $q$ can be obtained also for the case of $\varphi = 0$, when from (9) the following inequality is obtained

$$\cos^2 \psi < \frac{1}{2}(1 + \frac{ch\pi q - 1}{ch\pi q + 1}) = \frac{ch\pi q}{ch\pi q + 1} \qquad (11)$$

It is clear that when rising q to infinity, the region of $\psi$ for which collapse is possible monotonically extends to limits $0^0 < \psi < 180^0$, i.e. it corresponds to all $\psi$. The dependence of the energy E (9) on q for the different $\varphi$ is represented also on Fig.2

5. The Fig. 1 contains plots of the dependency of the upper and lower boundary threshold angle $\psi$ on $q$ for different values of $\varphi$ when the energy in (9) is positive, E>0.

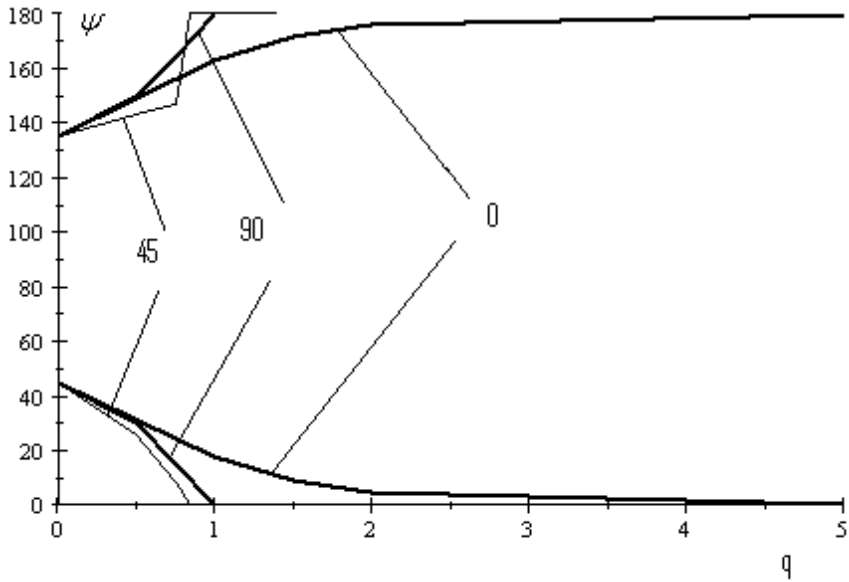

Fig. 1. Dependence of threshold angle $\psi$ on q for $\varphi$ =0, 45, and 90 degrees

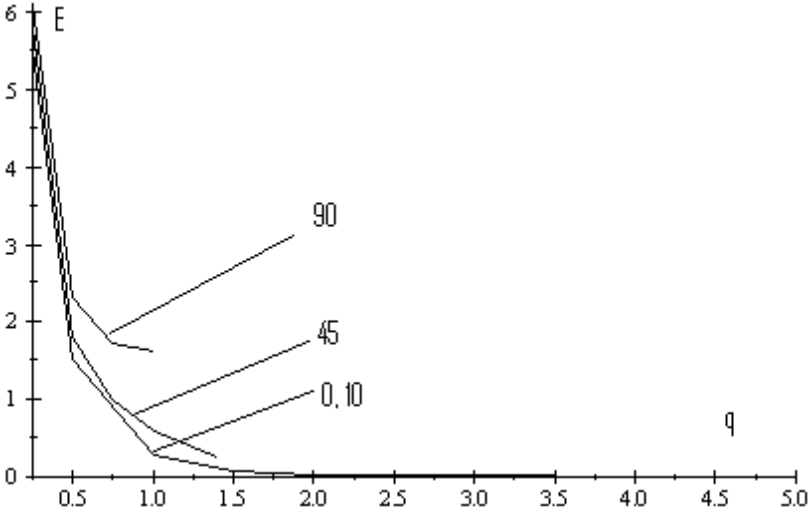

Fig. 2 Dependence of energy on q for $\varphi$ =0, 10, 45, and 90 degrees (the curves for $\varphi$ =0 and $\varphi$ =10 coincide)

Thus, from Fig.1, it follows the possibility for collapse realizing for all initial conditions, when q>2 or H<l/2. On the Fig2., from the other side, it is shown that the value of energy E may be smaller than 1/1000 for q>4 and the corresponding value of the collapse time $t_c \propto O(E^{-1/2})$ (see (6.1) and also (8)) for two PVD becomes too large for real observation in any experiment. The last restriction on q gives the interval 2<q<4 or $l(0)/4 < H < l(0)/2$ for the restriction on H (if the value of initial inter-particle distance $l(0) \approx n^{-1/3}$ is fixed by the concentration $n$ of particles in the unit of volume of the fluid), when it takes place real attraction for two particles (PVD), that corresponds to the known experimental results [1].

## *Conclusion and Discussion*

1. Here we have restricted our consideration by the particular case of two particles (PVD) dynamics only for illustrating the possibility of the hydrodynamic mechanism of the observed convergence of like-charged spheres if the ratio of their initial distance $l(0)$ to the layer depth $H$ have the restrictions from above and below (see Fig.1 and Fig.2).

From the other side, inferred in the present paper equations (2), (3) describing PVD dynamics in the bounded fluid layer allow considering a wide class of the hydrodynamics connected problems for any PVD number $N \geq 2$.

It is interesting to consider also PVD dynamics in the case of other geometry, e.g., in the pipe, including the case of the presence of the flow in the pipe. The latter might be used for modeling of charged erythrocyte structure forming when they move in the blood stream.

2. At the end of the section and paper, let us note also the possibility of using the conclusions obtained above (and in [16] on collapse condition for two particles-PVD interaction) in relation with well known investigation [23-25] of the process of scattering of two finite size vortex dipoles, which is considered in the limit of an unbounded fluid layer (when the parameter q=l(0)/H is equal to zero). As it is shown in [16] and in the theory presented above for q=0 (see (6.1) or (6), (10) when l/H=0), collapse of the two PVD can be realized only under condition: $|\cos\psi| < \sqrt{2}/2$, i.e. for the angles $\psi$ from the interval $45^0 < \psi < 135^0$. The value of the angle $\psi(0)$ (the angle between the direction of initial Lamb's impulse vector $\vec{\gamma}(0)$ and the vector $\vec{l}(0)$ connecting the two PVD at t=0) is related with the invariant value of the angular momentum by the following relation $I = \gamma l \sin\psi = \gamma\delta$. Here, parameter $\delta = l\sin\psi$ is impact parameter of scattering of the two PVD's with contrary directed Lamb impulse vectors $\vec{\gamma}$ (which are equal in modulus). It is found out here and in [16] that collapse of the two PVD is possible only under condition $\delta(0)/l(0) = \sin\psi > \delta_{th} = \sqrt{2}/2 \approx 0.707$. Let us note that in the process of collapsing, two particles (PVD) converge to one point in finite time and stay in the point for infinite time $T_{res} \to \infty$, forming the bound or long-live resonant states, analogue of which is observed in the numerical scattering data [23 - 25] for two finite vortex dipoles or two point vortex pairs the collision dynamics. Each of these vortex dipoles (or point vortex pairs) have the Lamb impulse $P_j^{(\alpha)} = k_\alpha \varepsilon_{ij} a_i^{(\alpha)}; \varepsilon_{11} = \varepsilon_{22} = 0, \varepsilon_{12} = -\varepsilon_{21} = 1; i = 1,2; j = 1,2; \alpha = 1,2$, where the two-dimensional vector $a_i^{(\alpha)}$ of the finite sized vortex dipole number $\alpha$ is directed from the point vortex with intensity $-k_\alpha$ to the other one constituting the vortex dipole with intensity $k_\alpha$ [23]. There exists the following relation with PVD Lamb's impulse: $\gamma_j^{(\alpha)} = \lim_{k_\alpha \to \infty; |a_i^{(\alpha)}| \to 0} P_j^{(\alpha)} = k_\alpha \varepsilon_{ij} a_i^{(\alpha)}$. In [23], for the same-sized and nearly the same intensity vortex dipoles (for $|\vec{a^{(1)}}| = |\vec{a^{(2)}}| = a; k_1 = k, k_2 = 1.1k$), it is stated that only for the impact parameter exceeding the critical value $\delta_c = \delta(0)/a \cong 0.785$, it is possible realization of the bound resonant state of the two vortex dipoles, when corresponding time of life for such resonant state abruptly increases [23]. In [23], it is noted that the resonant state effect is observed only when the distance between the dipoles centers l(0) has the same order of value as the own size of the vortex dipoles a(0). That is why, the threshold dimensionless value of the impact scattering parameter obtained in [23] may be equivalently defined in the nearly the same form $\delta_{th} = \delta(0)/l(0) \approx \delta_c \cong 0.785$, that corresponds even quantitatively to the mentioned above condition of the two, entirely identical by the Lamb impulse value, PVD collapse. Only for the relation l(0)=1.11a(0) we have also the strong equality $\delta_{th} = \delta_c$ between this two different determinations of the threshold impact parameter. Before appearance of the paper [24], the result obtained in [23] was interpreted on the base of the concept of the possibility of temporary (on the period of time with length $T_{res}$) formation of the vortex pairs with non-zero total intensity when converging of two finite-sized vortex dipoles on the distance of the order of their own initial size a(0). However, already in [24] and then in [25], there was stated an opportunity of the similar effect of forming of the bound states of dipoles also when scattering of two vortex dipoles with equal intensities $k_1 = k_2$. Nevertheless, when interpreting this fact, a key role in [24, 25] is also, as in [23], given to the finiteness of the sizes of the colliding vortex dipoles and, in particular, to the possibility of their different initial sizes when $a_1 \neq a_2$ and even $a_1 \gg a_2$ [24].

Basing on the noted above compliance with results [23–25] of the collapse condition of the two particles (PVD) for each of which the very zero individual space size is characteristic, we can give another qualitative interpretation to the results of the numerical investigation of the process of the finite size vortex dipoles scattering. Actually, according to the formula (14) from [16] (see also solution (6.1) for q=l(0)/H=0), when realizing the collapse of the two PVD in finite time $t = t_c$, they converge into the common point, and for $t > t_c$ already the solution becomes complex and has not physical sense. Therefore, we can guess that for the observed in the numerical experiments [23- 25] resonant states effects for the colliding of the two finite vortex dipoles, the main role, as for the scattering of the two PVD, is not played by their initial sizes. Note that as it is shown above, the role of the control dimensionless parameter is most suiting for the invariant value of the angular momentum *I* of the dipole vortex system (scaled by the product of the initial distance and Lamb impulse that corresponds to the $\sin\psi$), characterizing symmetry of the vortex dipole system. More over, from the obtained above and in [16] conditions of collapse of two PVD, it follows that for the control symmetry parameter $\psi$ there exists the restriction not only from below (for the unbounded fluid when q=0, it is $\psi > 45^0$), but also from above (when q=0, it is $\psi < 135^0$, see Fig. 1). That was not paid attention when numerical experiments results in [23-25] were provided. In [16] it is also consider the condition of collapse for two PVD in three dimensional case, when energy of PVD interaction is positive only for $1 > 3\cos^2\psi$ if $54.7^0 < \psi < 144.7^0$ in the unbounded fluid. Thus our consideration of PVD in two dimensional case gives the estimate from below for the upper limit of collapse condition because 135<144.7. It is also interesting to consider the more exact theory for three dimensional PVD in finite layer of fluid as we do here for two dimensional PVD.

Note also, that in [26], results of [23 – 25] are mentioned in the context of two different approaches to consideration of the finite vortex dipoles. In one of them, it is assumed invariance of the own sizes of each of the dipoles in spite of their interaction. Such invariance, strictly speaking, may take place either for the limit of PVD (that in [26] is considered but not on the base of an exact weak solution of hydromechanics equations as it was made in [16] for the case of two- and three-dimensional fluid), or for the vortex dipoles on the sphere when considering only invariant case of diameter-conjugated point vortexes having different signs and equal by module intensities (see [27]). From the other side in [16], where the very works [23 – 25] are mentioned, it is not excluded an opportunity of changing own sizes of the vortex dipoles in the process of their interaction. Obtained in the present work conclusion points that also for the processes of formation of the vortex dipole resonant states, considered in [23 – 25], actually, it is found acceptable also to assume invariance of the initial own sizes of the vortex dipoles when defining the conditions of formation of long-live vortex resonant state.

Thus, in our paper we give the new generalization of the Roberts - Chefranov equations (like they are named first in [31]) [16, 28-31] for the case of finite in one size fluid layer. This gives the possibility for wide application of the PVD- Particles method in hydrodynamics, plasma physics and biophysics. We show also that the symmetry of the PVD system is the real control parameter, but not the size of this vortex structure.

## Appendix

1. For a fluid layer of the finite depth $H$, it is not difficult using (1), to get (see in [21] formulas 5.1.25(3) and 5.1.25(4)) the following representation for the velocity field potential created by a system of the $N$ PVD :

$$\Phi = \sum_{\alpha=1}^{N}\Phi_{\alpha}, \Phi_{\alpha} = S + S(\gamma_y^\alpha \to -\gamma_y^\alpha; y_\alpha \to H - y_\alpha)$$

$$S = -\frac{1}{2\pi}\sum_{n=-\infty}^{\infty}\left[\frac{\gamma_x^\alpha(x-x_\alpha)+\gamma_y^\alpha(y-y_\alpha-2Hn)}{(x-x_\alpha)^2+(y-y_\alpha-2Hn)^2}\right] = -\frac{\gamma_x^\alpha b+\gamma_y^\alpha a}{4\pi H}S_1 + \frac{\gamma_y^\alpha}{4\pi H}S_2, \qquad (A.1)$$

$$S_1 = \sum_{n=-\infty}^{\infty}\frac{1}{(b^2+(n-a)^2)} = \frac{\pi sh2\pi b}{b(ch2\pi b - \cos 2\pi a)}, b = (x-x_\alpha)/2H, a = (y-y_\alpha)/2H$$

$$S_2 = \sum_{n=-\infty}^{\infty}\frac{n}{(b^2+(n-a)^2)} = 4a\sum_{n=1}^{\infty}\frac{n^2}{(n^2+b^2-a^2)^2+4a^2b^2} = \frac{\pi(ash2\pi b - b\sin 2\pi a)}{b(ch2\pi b - \cos 2\pi a)}$$

Similarly, we get representation for the stream function of the PVD system and the complex potential (2).

2. The system (3) can be also represented in the form:

$$dx_\beta/dt = \mathrm{Re}\,\frac{\pi}{8H^2}\sum_{\alpha=1}^{N}\left[\frac{\Gamma_\alpha}{sh^2(\frac{\pi(z_\beta-z_\alpha)}{2H})} - \frac{\Gamma_\alpha^*}{ch^2(\frac{\pi(z_\beta-z_\alpha^*)}{2H})}\right] = \mathrm{Re}\,F$$

$$dy_\beta/dt = \mathrm{Re}\,iF\,, \qquad (A.2)$$

$$d\gamma_x^\beta / dt = \operatorname{Re} F_1 = \operatorname{Re} \frac{\pi^2 \Gamma_\beta}{4H^3} \sum_{\alpha=1}^{N} \left[ \frac{\Gamma_\alpha cth \dfrac{\pi(z_\beta - z_\alpha)}{2H}}{ch \dfrac{\pi(z_\beta - z_\alpha)}{H} - 1} - \frac{\Gamma_\alpha^* th \dfrac{\pi(z_\beta - z_\alpha^*)}{2H}}{ch \dfrac{\pi(z_\beta - z_\alpha^*)}{H} + 1} \right]$$

$$d\gamma_y^\beta / dt = \operatorname{Re} i F_1,$$

where Re denotes the operation of extraction of the real part of the expression at the right; and index $\beta = 1,...,N$.

3. In the case when in (3) and (A.2), the PVD number is equal to two, N=2, we get in the general case a system of equations following from (3) and (A.2) and having the following form (for $R_y = (y_1 + y_2)/2; l_x = x_1 - x_2, l_y = y_1 - y_2$):.

$$dl_x / dt = \frac{\pi}{4H^2} \left[ \frac{a_0(\gamma_{2x} - \gamma_{1x}) + b_0(\gamma_{2y} - \gamma_{1y})}{a_0^2 + b_0^2} - \frac{a_1(\gamma_{2x} - \gamma_{1x}) - b_1(\gamma_{1y} + \gamma_{2y})}{a_1^2 + b_1^2} \right]$$

$$dl_y / dt = \frac{\pi}{4H^2} \left[ \frac{-a_0(\gamma_{2y} - \gamma_{1y}) + b_0(\gamma_{2x} - \gamma_{1x})}{a_0^2 + b_0^2} - \frac{a_1(\gamma_{2y} - \gamma_{1y}) + b_1(\gamma_{1x} + \gamma_{2x})}{a_1^2 + b_1^2} \right] \quad (A.3)$$

In (A.3), we use the following notations:

$$\gamma_x^\beta = \gamma_{\beta x}, \gamma_y^\beta = \gamma_{\beta y}, \beta = 1,2; a_0 = ch\frac{\pi l_x}{H}\cos\frac{\pi l_y}{H} - 1, b_0 = sh\frac{\pi l_x}{H}\sin\frac{\pi l_y}{H},$$

$$a_0^2 + b_0^2 = (ch\frac{\pi l_x}{H} - \cos\frac{\pi l_y}{H})^2; a_1 = 1 + ch\frac{\pi l_x}{H}\cos\frac{\pi R_y}{H}, b_1 = sh\frac{\pi l_x}{H}\sin\frac{\pi R_y}{H}.$$

And for the Lamb impulse components, we have

$$d\gamma_{1x} / dt = \frac{\pi^2}{4H^3} \left[ \frac{a_{01}A_1 sh\dfrac{\pi l_x}{H} + b_{01}A_2 \sin\dfrac{\pi l_y}{H}}{c(a_0^2 + b_0^2)} - \frac{a_{11}A_3 sh\dfrac{\pi l_x}{H} + b_{11}A_4 \sin\dfrac{2\pi R_y}{H}}{c_1(a_1^2 + b_1^2)} \right],$$

$$a_{01} = a_0 - \sin^2\frac{\pi l_y}{H}, b_{01} = a_0 + sh^2\frac{\pi l_x}{H}, a_{11} = a_1 + \sin^2\frac{2\pi R_y}{H}, b_{11} = a_1 - sh^2\frac{\pi l_x}{H}, \quad (A.4)$$

$$c_1 = ch\frac{\pi l_x}{H} + \cos\frac{2\pi R_y}{H}, c = ch\frac{\pi l_x}{H} - \cos\frac{\pi l_y}{H}; A_1 = \gamma_{1x}\gamma_{2x} - \gamma_{1y}\gamma_{2y},$$

$$A_2 = \gamma_{2x}\gamma_{1y} + \gamma_{2y}\gamma_{1x}, A_3 = \gamma_{1x}\gamma_{2x} + \gamma_{1y}\gamma_{2y}, A_4 = \gamma_{1x}\gamma_{2y} - \gamma_{1y}\gamma_{2x};$$

$$d\gamma_{2x} / dt = -d\gamma_{1x} / dt.$$

$$d\gamma_{1y} / dt = \frac{\pi^2}{4H^3} \left[ \frac{-a_{01}A_2 sh\dfrac{\pi l_x}{H} + b_{01}A_1 \sin\dfrac{\pi l_y}{H}}{c(a_0^2 + b_0^2)} - \frac{a_{11}A_4 sh\dfrac{\pi l_x}{H} - b_{11}A_3 \sin\dfrac{2\pi R_y}{H}}{c_1(a_1^2 + b_1^2)} \right], \quad (A.5)$$

$$d(\gamma_{1y}+\gamma_{2y})/dt = -\frac{\pi^2(a_{11}A_4 sh\frac{\pi l_x}{H} - b_{11}A_3 \sin\frac{2\pi R_y}{H})}{2H^3 c_1(a_1^2+b_1^2)}.$$

For defining the evolution with time of the value $R_y$, we have the following equation

$$dR_y/dt = \frac{\pi}{8H^2}\left[\frac{-a_0(\gamma_{1y}+\gamma_{2y})+b_0(\gamma_{1x}+\gamma_{2x})}{a_0^2+b_0^2} - \frac{a_1(\gamma_{1y}+\gamma_{2y})+b_1(\gamma_{1x}+\gamma_{2x})}{a_1^2+b_1^2}\right] \qquad (A.6)$$

From (A.4)-(A.6), it follows that for the initial data with $R_y = 0, \gamma_{1x}+\gamma_{2x}=0, \gamma_{1y}+\gamma_{2y}=0$, the both components of the both PVD summed Lamb impulse are invariant with respect to time and the both are equal to zero (when $\vec{\gamma}_2 = \vec{\gamma} = -\vec{\gamma}_1$). In this case the energy of two PVD takes form (9) and the corresponding dynamic system have the next presentation:

$$dl_x/dt = \frac{\pi}{2H^2}\left[\frac{\gamma_y b_0}{(a_0^2+b_0^2)^2} - \gamma_x(\frac{1}{ch(l_x\pi/H)+1} - \frac{a_0}{(a_0^2+b_0^2)^2})\right];$$

$$dl_y/dt = \frac{\pi}{2H^2}\left[\frac{\gamma_x b_0}{(a_0^2+b_0^2)^2} - \gamma_y(\frac{1}{ch(l_x\pi/H)+1} + \frac{a_0}{(a_0^2+b_0^2)^2})\right]; \qquad (A.7)$$

$$d\gamma_x/dt = \frac{\pi^2}{4H^3}\left[\frac{2\gamma_x\gamma_y(a_0+sh^2(l_x\pi/H))\sin(l_y\pi/H)+(\gamma_x^2-\gamma_y^2)(a_0-\sin^2(l_y\pi/H))sh(l_x\pi/H)}{(a_0^2+b_0^2)^{3/2}} - K_0\right],$$

$$K_0 = \frac{(\gamma_x^2+\gamma_y^2)sh(l_x\pi/H)}{(ch(l_x\pi/H)+1)^2};$$

$$d\gamma_y/dt = -\frac{\pi^2}{4H^3}\left[\frac{2\gamma_x\gamma_y(a_0-\sin^2(l_y\pi/H))sh(l_x\pi/H)-(\gamma_x^2-\gamma_y^2)(a_0+sh^2(l_x\pi/H))\sin(l_y\pi/H)}{(a_0^2+b_0^2)^{3/2}}\right].$$

4. Let's obtain the weak solution of two – dimensional hydrodynamic equations for the case of unbounded fluid, like it do for the three dimensional case in [16]. The same derivation was also made early by V. M. Gryanik (in private communication, see also [16] about this).

The Helmholtz equation for the scalar vortex field $\omega$ in the two – dimensional case (for the ideal incompressible fluid) have the form:

$$\frac{\partial\omega}{\partial t} + u_l\frac{\partial\omega}{\partial x_l} = 0, \qquad (A.8)$$

Substituting the expression for $\omega$ from (1.1) (see the first paragraph of the paper) into (A.8) and multiplying both sides of (A.8) by an arbitrary smooth function $\varphi$, we can carry out the integration over the entire space, including the region in which the singularity points of PVD localization in (1.1) are concentrated. From (A.8) and (1.1) we find:

$$\sum_{m=1}^{N}\int d^2x\delta(\vec{x}-\vec{x}_m)\left[-\varepsilon_{lp}\frac{d\gamma_p^m}{dt}\frac{\partial\varphi}{\partial x_l} + \varepsilon_{jk}\gamma_k^m(\frac{\partial u_l}{\partial x_j}\frac{\partial\varphi}{\partial x_l} + \frac{\partial^2\varphi}{\partial x_j\partial x_l}(u_l - \frac{dx_l^m}{dt}))\right] = 0 \quad (A.9)$$

where m=1, 2, …, N. From (A.9) we have the finite dimensional Hamiltonian system of ordinary differential equations, which represent the weak solution of (1):

$$\frac{dx_l^m}{dt} = u_l(\vec{x_m}), \qquad (A.10)$$

$$\frac{d\gamma_l^m}{dt} = -\gamma_k^m \frac{\partial u_k(\vec{x_m})}{\partial x_l^m}, \qquad (A.11)$$

where for obtain (A.11) we use the equation $\varepsilon_{rp}\frac{d\gamma_p^m}{dt} = \varepsilon_{jk}\gamma_k^m \frac{\partial u_r}{\partial x_j}$ (after multiplying it on $\varepsilon_{rl}$ we obtain (A.11)).